\documentclass[showpacs,showkeys,preprint]{revtex4}
\usepackage{amsmath,epic}

\preprint{SNUTP 04-011}

\begin{document}
\title{Spectra of Heterotic Strings on Orbifolds}
\author{Kang-Sin Choi}\email{ugha@phya.snu.ac.kr}
\affiliation{School of Physics and Center for Theoretical Physics,
Seoul National University, Seoul 151-747, Korea }

\begin{abstract}
We obtain the spectrum of heterotic strings compactified on
orbifolds, focusing on its algebraic structure. Affine Lie algebra
provides its current algebra and representations. In particular, the
twisted spectrum and the Abelian charge are understood. A twisted
version of algebra is used in the homomorphism from the orbifold
action to the group action. The relation between the conformal
weight and the mass gives a useful rule.
\end{abstract}



\maketitle

\newcommand{\ad}{{\rm ad}}
\newcommand{\Z}{{\bf Z}}
\newcommand{\hwr}{|\mu\rangle}
\newcommand{\balpha}{\bar {\alpha}}
\newcommand{\1}{{\bf 1}}
\newcommand{\2}{{\bf 2}}
\newcommand{\3}{{\bf 3}}
\newcommand{\8}{{\bf 8}}
\newcommand{\Tr}{{\rm Tr}}
\newcommand{\ch}{{\rm ch}}
\newcommand{\g}{{\sf g}}
\newcommand{\h}{{\sf h}}
\newcommand{\T}{{\sf T}}

\section{Introduction}
Weakly coupled heterotic strings compactified on orbifolds
\cite{DHVW,IMNQ} have long garnered a great deal of attention
due to their several desirable features of low energy theory. They
contribute to the understanding of such things as gauge groups, matter spectra, number of
families, and small supersymmetries, to name a few. They are very predictive,
compared to, for example, field theoretic orbifold models, because
string theory places restrictions on possible spectra in the bulk and
at the fixed points.

In the process of building a model, the most severe obstacle is the
fact that there are too many possible string vacua. It is expected
that a realistic orbifold compactification includes more than one
Wilson line in order to have a sufficiently small group and number
of families \cite{INQWilson, IKNQ}. In distinguishing the gauge
groups and matter spectra, it is currently known that there are more
than $\sim 10^7$ possibilities in the simplest $T^6 / \Z_3$
orbifolds with two Wilson lines \cite{CMM, CHK}. However, such a
large estimate is due to a lack of understanding of symmetries. We
recently classified all the gauge groups and untwisted matter
spectra with the aid of some group theoretical methods and observed
that the number is dramatically lower \cite{CHK}.

However a difficulty still lies in understanding the twisted
sector spectrum. We know that the modular invariance condition
requires the twisted strings, characterized by the periodic
boundary condition up to automorphisms. The twisted strings form
another independent Hilbert space, whose structure is poorly
understood. We obtain the twisted string spectrum from the mass
shell condition of strings supplemented by GSO projection, but it
remains unclear how to determine and analyze its algebraic
structure. Again, the task of finding abelian generators was a
trial-and-error job.

In this paper, we seek such a structure with the aid of algebraic
tools. In this endeavor, the rich algebraic structure provided by
the affine Lie algebra proves to be useful. The vertex operator
construction \cite{FKS} teaches us that the heterotic string
spectrum is the representation of affine Lie algebra.

In particular, the twisted sector states form the representations of
a twisted version of affine Lie algebra, defined in the same manner
as twisting the physical states. When the twisting is inner
automorphism, with a suitable change of basis as we will see, the
algebra is isomorphic to the original one, making it easy to obtain
twisted representations \cite{Kac,FuSc}.

This also provides us a systematic way to obtain and classify groups
and spectra. We will see that the affine Lie algebra plays a crucial
role in understanding the Abelian group embedded in the non-Abelian
group as well. The related anomalous $U(1)$ is also an important
issue \cite{CKM,FIQS,KN}.

This paper focuses on  modular invariant theory and level one
algebra. However, the discussion still holds for those without
modular invariance or those with higher level algebras \cite{Di}.
Most of the mathematical facts used here can be found in Ref.
\cite{Kac,FuSc,GO}. We will follow convention on naming roots and
weights of \cite{FuSc}, which is also presented in the appendix.

\section{Affine Lie algebra}
Consider an algebra $\g$, whose generators satisfy the commutation
relation,
\begin{equation}
 [T^a_m,T^b_n]=if^{abc}T^c_{m+n} + m\delta_{m+n,0} \delta^{ab} K.
 \label{kk}
\end{equation}
The group indices $a,b,c$ run over the dimensions of the algebra
$d=\rm \dim \sf g$ and the `mode' indices $m,n$ are integers. This
is an extension of the simple Lie algebra $\bar \g$ for which
$m=n=0$. Without $K$ term, $T^a_m$ are understood as infinitesimal
generators of mapping $S^1 \to \g$, from those of simple Lie
algebra $T^a_0$. This procedure is known as an affinization, so
this algebra is called the {\em affine Lie algebra}, or the {\em
Kac--Moody algebra}. We use the overlined letter for the objects
of the simple Lie algebra.

We introduce two additional generators. One is the {\em central
element} $K$, commuting with all the generators
\begin{equation}
 [K,T_m^a]=0,
\end{equation}
which is made of a linear combination of the Cartan generators $H$
and turns out to be unique. The other is the {\em grade operator}
$D$ which has the following commutation relations
\begin{equation}
[D,T^a_m] = m T^a_m, \quad \left[D,K \right] = 0. \label{grade}
\end{equation}
We collectively denote the weights of $\sf g$ as the eigenvalues of
$(H_0,K,D)$. Inspecting the Killing form, the natural inner product
is
\begin{equation}
(\bar \lambda,k,n) \cdot (\bar \lambda',k',n')
 = \bar \lambda \cdot \bar \lambda' + kn' + k'n. \label{product}
\end{equation}
We have simple roots of $\g$ by extending those $\bar \alpha^i$ of
the simple Lie algebra $\bar {\g}$, with the highest root $\bar
\theta$,
\begin{eqnarray}
\alpha^i &=&(\bar \alpha^i,0,0),  \quad i=1,\cdots,r \\
\alpha^0 &=&(-\bar \theta,0,1)
\end{eqnarray}
where $r$ is the rank of $\bar \g$. The $\alpha^0$ will raise and
lower the eigenvalue $D$ because of the $(0,0,1)$ component.

It is useful to define the dual vector $\alpha^\vee = 2\alpha /
\alpha^2$. Constructing the Cartan matrix $A^{ij}=2 \alpha^i \cdot
\alpha^{j \vee}$, we obtain the Dynkin diagram of $\g$, which is
the same as the extended Dynkin diagram of $\bar \g$.

With the positive numbers $a_i$ known as the {\em dual Coxeter
labels}, we have the unique expansion $\bar \theta^\vee = \sum a_i
\bar \alpha^{i \vee}$ and $a_0 = 1$. The sum $g=a_0+\sum a_i$ is
called the {\em dual Coxeter number}. An example for $E_8$ and
$SO(8)$ will be given in Figs. 1 and 2. For a list of Dynkin
diagrams and Coxeter numbers, see Ref. \cite{FuSc}. We also define
fundamental weights $\Lambda_i$ as $\alpha^{i\vee} \cdot \Lambda_j =
\delta^i_j$, therefore
\begin{equation} \label{fundwghts}
\begin{split}
 \Lambda_i &= (\bar \Lambda,\frac12{a_i \bar \theta^2} ,0) \\
 \Lambda_0 &= (0,\frac12{\bar \theta^2},0)
\end{split} \end{equation}
and the Weyl vector $\rho=\sum_1^r \Lambda_i + g\Lambda_0$. The
fundamental weight has a relation with the inverse of the Cartan
matrix $\bar \Lambda_i \cdot \bar \Lambda_j = A^{-1}_{ij}$,
sometimes refered to as $\lq$quadratic form' in the literature. From
the definition of the Weyl vector, it follows that
\begin{equation}
 g \bar \theta^2 \delta^{ab}=f^{acd}f^{bcd}, \label{normalization}
\end{equation}
the quadratic Casimir for the adjoint representation. We will follow
convention when referring to the simple roots and the fundamental
weights of reference \cite{FuSc} and their explicit form can also be
found in the appendix.

By triangular decomposition, we separate generators into the
Cartan subalgebra and the ladder operators
\begin{eqnarray}
 && [H^i_m,H^j_n]=m \delta^{ij} \delta_{m+n,0} K,
 \label{extcartan} \\
 && [H^i_m,E^{\balpha}_n] = \balpha^i E^{\balpha}_{m+n},
 \label{raising} \\
 && [E^{\balpha}_m,E^{\bar \beta}_n] = \pm
    E^{\balpha+\bar \beta}_{m+n}\mbox{    for } \balpha
    +\bar{\beta} \in {\sf \bar g} \mbox{ root,} \label{ladder} \\
 && [E_n^{\balpha},E^{-\balpha}_{-n}]= \balpha \cdot
 H_0 + n K. \label{halpha}
\end{eqnarray}
Again we see that these are extended relations of the simple Lie
algebra. Note that the grade can also be raised and lowered by
generator $T_n^a$ with $n \ne 0$. For example, the Cartan subalgebra
does not mutually commute anymore, and by
$$ [H^i_n,H^j_{-n}]=\delta^{ij} n K, $$
each of which raise and lower the eigenvalue of $D$ by $nk'$, in
accord with the product (\ref{product}). We have the highest weight
module, whose highest weight vector is annihilated by all the
positive roots including the positive grade operator,
\begin{eqnarray}
 E^{\bar \alpha}_0 |\Lambda \rangle = 0, & \mbox{for all }
 \bar \alpha >0, \label{hgstst} \\
 T^a_n |\Lambda \rangle = 0, & \mbox{for all } n>0.
 \label{graderaise}
 \end{eqnarray}
The complete vectors in the module are obtained by a series of
lowering operations. Owing to the creation operator for $n$, such
a weight system is, generally speaking, infinite dimensional.

Here we are interested in a finite dimensional submodule. For any
weight $\lambda= (\bar \lambda,k',n)$ of a highest weight module,
\begin{equation}
  {2 \over \alpha^2} [E^{\bar \alpha}_n,E^{-\bar
  \alpha}_{-n}] |\lambda
  \rangle
 = {2 \over {\bar \alpha}^2} (\bar \alpha \cdot \bar \lambda+
nk') | \lambda \rangle
\end{equation}
from the relation (\ref{halpha}). From the $SU(2)$ representation
theory, for a finite dimensional module, all of the eigenvalues are
integers. So it follows from simple Lie algebra $\bar \g$ that each
eigenvalue in the sum is a separate integer. Applying to the highest
weight state, in the $\alpha^0=(-\bar \theta,0,1)$ direction,
\begin{equation}
 \left(-\bar \Lambda \cdot {\bar\theta}^\vee+ {2 \over
{\bar\theta}^2}k' \right)|\Lambda\rangle.
\end{equation}
Now the eigenvalues are nonnegative integers. Therefore, the
quantity $k \equiv 2k' / \bar \theta^2$, called the {\em level},
is a nonnegative integer.

We are interested in the level one algebra, therefore hereafter we
set $k=1$, although we explicitly bare the letter $k$ for extension.
For the highest weight $\bar \Lambda=\sum t_i \bar \Lambda_i$, $t_i$
is a nonnegative integer, by the definition of the fundamental
weight. Then the above relation leads to the so-called integrability
condition
\begin{equation}
 0 \le \sum_{i=1}^r a_i t_i \le k.
\end{equation}
The above relation naturally extended to affine Lie algebra in a
tidy form. From $\Lambda=\sum t_i \Lambda_i=(\bar \Lambda, \frac12
\bar \theta^2 (t_0+ \sum_{i=1}^r a_i t_i),0)$, the middle element
being the level, we have an equivalent relation
\begin{equation}
 k = t_0 + \sum_{i=1}^r a_i t_i. \label{integrability}
\end{equation}
It is noted that for the level one algebra only a few can satisfy
this condition. For the $SU(n)$ algebra, the dual Coxeter label
corresponding to a fundamental weight is always $1$; thus every
representation $\Lambda_i$, having dimension $\binom{n}{i}$, is
possible. However, the other groups have $a_i=1$ only for the
outer most nodes of the Dynkin diagram. For example, for $E_6$,
$\Lambda_1 ({\bf 27})$ and $\Lambda_6 (\overline{\bf 27})$ have
$a_i=1$ thus can satisfy this condition.

A corollary of this observation is that, in the $k=1$ case the
adjoint representation, which can become an adjoint Higgs field
used to break typical $SU(5)$ and $SO(10)$ unified theory, cannot
satisfy this condition. The weight vector for the adjoint
representation of $SU(n)$ is $\Lambda_1+\Lambda_{n-1}$, hence it
has the sum of the dual Coxeter labels greater than 1:
$a_1+a_{n-1}=2 > 1=k$. Of course for the higher level algebra, we
can have such adjoint representation \cite{Di}.

\section{Heterotic string}

\subsection{Current algebra}

Heterotic string theory has closed strings with one worldsheet
supersymmetry on the right movers. For the left movers, in which
we are interested in this section, on top of the ten spacetime
bosonic degrees of freedom, 16 extra bosons (or 32 fermions) are
needed to cancel the conformal anomaly. The modular invariant
theory allows the gauge group $SO(32)$ or $E_8 \times E_8$
\cite{GHMR}.

One way to denote the group degrees of freedom is by state vectors
$|p \rangle$. In the bosonic description, they represent charges in
given directions, thus spanning the root space. In the low energy
limit, massive modes of the order of the string scale are decoupled.
Resorting to the mass shell condition (in unit $\alpha'=1$),
\begin{equation}
 { M_{\rm L}^2 \over 4 } = { p^2 \over 2 } + \tilde N - 1
 \label{mommass}
\end{equation}
and the modular invariance, one notes that the zero mode vector
$p$ lives in the even and self-dual lattice, which turns out to be
the $SO(32)$ or $E_8 \times E_8$ lattice. The Cartan generators
are provided by oscillators $\tilde \alpha_{-1}^i |0 \rangle$.
Combined with massless right movers, they constitute the fields
for supergravity coupled with this gauge group.

Despite their clear spectra, the role of massive modes is not
transparent. To have a better understanding, we look at another
equivalent description. In conformal field theory, every state has
one to one correspondence with a local `vertex operator'. They
look like $T^i(z) \psi(\bar z) e^{i k \cdot X}$ with $T^i(z)$
being \cite{FKS}
\begin{align}
 H^i(z) &=  \partial_z X^i, \label{u1vertex} \\
 E^\alpha(z) &= \alpha^i E^i = c_\alpha :e^{i \alpha \cdot X}: \label{laddervertex},
\end{align}
where $c_\alpha$ is the two-cocycle determining the sign of the
commutator as in (\ref{ladder}) and colons denoting the
conventional normal ordering. The operator product expansion
(whose expectation value is a two-point correlation function)
between two currents is, as $z\to 0$,
\begin{equation}
 T^a(z) T^b(0) \sim {k' \delta^{ab} \over z^2} + {ic^{abc} \over z } T^c(0).
 \label{ope}
\end{equation}
With the mode expansion $T^a(z)= \sum T^a_n z^{-n-1}$, and
identifying $c^{abc}\equiv f^{abc}$ they satisfy the commutation
relation of the affine Lie algebra
(\ref{kk},\ref{extcartan}--\ref{halpha}). This normalization also
fixes the level $k = 2 k'/\bar \theta^2 =1$. In the heterotic
string theory embedding, we have only the level one vertex
operators. However, we can make higher level algebra by embedding
it into the product of level one algebras, for example \cite{Di}.

The two dimensional conformal symmetry, which the string theory
possesses, is realized by the analytic transformation $z'=f(z)$
(i.e. independent of $\bar z$) \cite{FMS}. (Since the all the
currents are from the left mover, we will focus on analytic
currents in $z$ for the present purpose.) Under it, the vertex
operator $O(z)$ transforms as
\begin{equation} \label{conftransf}
 O(z') = \left({\partial z' \over \partial z} \right)^{-h} O(z),
\end{equation}
where we define $h$ as {\em conformal weight}. If an operator
transforms definitely under (\ref{conftransf}), we call it {\em
primary operator}. We can check that the currents
(\ref{u1vertex},\ref{laddervertex}) are primary operators of
conformal weight one. The mapping from a state to a vertex
operator corresponds to shrinking (conformal transformation) the
`in' and `out' states of a cylindrical Feynman diagram into
points. All the quantum number is kept and we will see that
especially the mass is converted into the conformal weight.

By Sugawara construction \cite{Su,FMS}, we construct a worldsheet
energy--momentum tensor ${\T}(z)$ from the generators of the
algebra $\g$,
\begin{equation}
 {\T}(z)={1 \over \beta} \sum_{a=1}^d :T^a(z) T^a(z):.
\end{equation}
This is the generator of the conformal symmetry
(\ref{conftransf}). The normalization $\beta$ will be fixed by
requiring for $T^a(z)$'s to transform as the primary fields of
conformal weight one \cite{FMS},
\begin{equation}
 {\T}(z) T^a (0) \sim {T^a(0) \over z^2} + {\partial T^a(0) \over
 z}.
\end{equation}
which implies
\begin{equation}
 [L_m,T^a_{-n}]=n T^a_{m-n}. \label{hamiltonian}
\end{equation}
Doing Fourier expansion $\T(z)=\sum L_n z^{-n-2}$, we have its
components
\begin{equation}
 L_n = {1 \over \bar \beta}
 \sum_{m\in \Z} \sum_{a=1}^{d} : T^a_{m+n} T^a_{-m} :.
 \label{sugawara}
\end{equation}
Now we have another way to express conformal vacuum
(\ref{graderaise}),
\begin{equation} \label{confvac}
 L_n | \Lambda \rangle = 0, \quad n >0.
\end{equation}
Acting $L_{-1}$ we have
$$ L_{-1} |\Lambda \rangle = \frac 2 \beta T^a_{-1} T^a_0 |\Lambda \rangle. $$
Using (\ref{kk}) and (\ref{hamiltonian}), we get
\begin{equation} \begin{split}
 T^a_0 |\Lambda \rangle &= \frac 2 \beta (i f^{bac}T_0^c + k'
 \delta^{ab}) T^a_0 |\Lambda \rangle \\
 &= \frac 1 \beta (i f^{bac} i f^{dca} T^d_0 + 2k'T^b_0) |\Lambda \rangle \\
 &= \frac 1 \beta (\bar \theta^2 g + \bar \theta^2 k) T^b_0
 |\Lambda\rangle,
\end{split} \end{equation}
where in the last line we used the property of the quadratic
Casimir and $g$ (\ref{normalization}), and definition of the level
$k$. Therefore we have the normalization $\beta = \bar \theta^2 (k
+ g) $. Finally we have the Virasoro algebra,
\begin{equation}
 [L_m,L_n] = (m-n)L_{m+n}+\frac {c^{\g}} 2 (m^3-m)\delta_{m+n,0},
\end{equation}
with the {\em conformal anomaly},
\begin{equation}
 c^{\g}={k d \over k + g \bar \theta^2/2 }. \label{confanomaly}
\end{equation}
We usually normalize $\bar \theta^2 = 2$. In addition, for $k=1$,
with a simple relation $g+1=d/r$ \cite{FuSc}, it coincides with
the rank $ c^{\g} = r$.

\subsection{Finding the state}

We want to find the physical state (\ref{confvac}),
$$
 L_n | \Lambda \rangle = 0, \quad n >0.
$$
Then, the worldsheet Hamiltonian is $L_0$ which gives rise to the
mass squared operator of string states and has a lower bound. Note
the commutator (\ref{hamiltonian})
$$
 [L_m,T^a_{-n}]=n T^a_{m-n}.
$$
Putting $m=0$, we see that $|\Lambda\rangle$ is an eigenstate of
$L_0$ and $D$ up to an additive constant. We use additive
normalization $L_0 |1\rangle = D |1\rangle = 0$, where $|1\rangle$
is an uncharged physical vacuum (the ground state of a unitary
conformal field theory). As a corollary, the gauge boson generators
should be zero modes since they should commute with the worldsheet
Hamiltonian $L_0$, and not change the mass. By (\ref{confvac}), the
eigenvalue $h_\Lambda$ is positive definite and minimal. We have
\begin{equation}
 \frac {M_{\rm L}^2} 4 = h_\Lambda - \frac {c} {24} =  n.
 \label{massshell}
\end{equation}
instead of (\ref{mommass}). The oscillator part $\tilde N$ is going
to be contained in $h_\Lambda$ later. Note that, in the additive
normalization, all the fields of given conformal field theory
contribute to the conformal anomaly $c$. We have not specified the
whole, i.e. spacetime degrees of freedom, as we will see in
(\ref{spcanomaly}), so that $c=c^{\sf g}+c^{\sf s}$.

The eigenvalue of $L_0$, or the conformal weight $h_\Lambda$ is
obtained if we apply (\ref{sugawara})
\begin{equation} \begin{split}
 L_0|\Lambda \rangle &= {1 \over \bar \theta^2 (k+g)} \sum_{m \in
 \Z}\sum_{a=1}^d :T^a_{m+n} T^a_{-n} : |\Lambda\rangle \\
 &= {1 \over \bar \theta^2 (k+g)} \sum_{a=1}^d T_0^a T_0^a |\Lambda
 \rangle = {C_r \over \bar \theta^2 (k+g)} |\Lambda \rangle.
\end{split} \end{equation}
Here $C_r$ is the quadratic Casimir in the representation of
$\Lambda$ and explicitly expressed in terms of fundamental weights,
using the property of the Weyl vector \cite{FuSc}. Therefore we
obtain so-called the Freudenthal--de Vries strange formula
\cite{Ka},
\begin{equation}
 h_\Lambda = {\Lambda \cdot ( \Lambda +2 \rho ) \over \bar
\theta^2(k+g)}.
 \label{confwght}
\end{equation}
Explicitly with $\Lambda=\sum_1^r t_i \Lambda_i$ and $\Lambda_i
\cdot \Lambda_0 = 0$, we have
\begin{equation}
 h_\Lambda = \frac 1{\bar \theta^2(k+g)} \sum_{i,j=1}^r (t_i+2) t_j
A_{ij}^{-1}.
\end{equation}
In case the state $|\Lambda\rangle $ consists of a single
fundamental weight $\Lambda_i$, (we will consider more general case
soon) it reduces to
\begin{equation}\label{length}
 h_{\Lambda_i} =  \frac12 \Lambda_i^2 = A^{-1}_{ii} \quad
 \mbox{(no summation of $i$),}
\end{equation}
using the property of the Coxeter labels \cite{FuSc}.

The problem of finding states satisfying the mass shell condition
(\ref{massshell}) is converted to the finding $h_{\Lambda}$
satisfying it. Now the task is to seek the explicit form of the
highest weight vector $\Lambda$ by reading off the inverse Cartan
matrix $A^{-1}$. This is convenient because $A^{-1}$ is uniquely
defined, independent of the basis vector one uses. The
corresponding highest weight vector is obtained from the following
relation;
\begin{equation} \label{hgstwgtvec}
 p = \bar \Lambda_i = \sum_j A^{-1}_{ij} \bar \alpha^{j \vee}
\end{equation}
where we carefully use $\bar \alpha^{j \vee}$ as a simple (dual)
root of {\em original} $E_8 \times E_8$, in which the algebra is
embedded. The complete weight vectors are obtained by successive
lowering with the aid of the Cartan matrix.

In general the state is charged under more than one simple (and
Abelian) algebra. Typically, it is the fixed point algebra
resulting from the breaking of $E_8 \times E_8$ or $SO(32)$.
Denote the whole algebra as $\bigoplus \h$ and the conformal
weight of each simple algebra as $h_{\Lambda^{\h}}$. Since the
conformal weight is additive, we may replace the $h_\Lambda$ in
(\ref{length}) as the sum over the whole algebra,
\begin{equation} \label{sumh}
 h_\Lambda = \sum h_{\Lambda^{\h}}.
\end{equation}
Also because such algebras are disconnected, we can easily write the
weight vector $p$ in (\ref{hgstwgtvec}) as also the sum of the
weight vectors of each algebra
\begin{equation} \label{sump}
 p = \sum \bar \Lambda^{\h}_i.
\end{equation}
We will see shortly that this still holds for the Abelian groups.

Now we can interpret massive modes in terms of algebraic operators.
The grade raising generators $H^i_{-n}$ and $E^\alpha_{-n} (n>0)$
correspond to $\alpha^i_{-n-1} |0\rangle$ and $ |p\rangle $ with
$p^2/2=n+1$, respectively, by (\ref{massshell}). For explicit
spectra with $n=1$, see \cite{GSW}. With more compact dimensions, we
may generalize this idea to the case of Narain compactification
\cite{Narain}.

\section{Twisting algebra}

\subsection{Twisted algebra}

Let us define shift vector $v$ by which the action on states is
\begin{equation} \label{shift}
  \omega = \exp  (2 \pi i v \cdot \ad_H )
\end{equation}
where ad is the commutation `adjoint' operation $\ad_A B = [A,B]$.
Noting that $[H^i,E^{\bar \alpha}] = \bar \alpha^i E^{\bar
\alpha}$, we have
$$ \omega |p \rangle = \exp (2\pi i v \cdot p) |p \rangle. $$
The order of automorphism $l$ is the minimal integer such that
$\omega^l=1$, so that $l v$ belongs to the (dual) weight lattice. It
is known that every inner automorphism of finite order can be
represented by such a shift \cite{FuSc}. For the case of $E_8$, the
Dynkin diagram possesses no symmetry, thus the only automorphism is
inner automorphism \footnote{The full $E_8 \times E_8$ does have
relevant outer automorphism can be used to break the group
\cite{dB}. Although SO(32) does possesses an outer automorphism, the
complex conjugation, it is not relevant since there is no effect on
the adjoints.}. Such an automorphism naturally comes from the
twisted string current
\begin{equation}
 T^a(e^{2\pi i} z) = e^{2\pi i \eta^a} T^a(z)
\end{equation}
with $l \eta \in \Z$. The same terminology is used for the
algebra. They satisfy the twisted algebra
\begin{equation}
[T^a_{m+\eta^a},T^b_{n+\eta^b}]=i f^{abc}T^c_{m+n+\eta^a+\eta^b} +
   (m+\eta^a) \delta_{m+n+\eta^a+\eta^b,0} \delta^{ab} K.
\end{equation}
Under the triangular decomposition (\ref{extcartan}-\ref{halpha}),
the twisting (\ref{shift}) leaves the Cartan generators invariant.
Thus, we still have the integer moded generators. This should be
true also for raising generators by some redefinition. To absorb the
twist in the $\bar \alpha$ direction, we define
\begin{equation}
 \tilde E^{\bar \alpha}_n = E^{\bar \alpha}_{
n+{\bar \alpha} \cdot v}, \label{root}
\end{equation}
To compensate for this, we require
\begin{eqnarray}
&& \tilde H^i_n = H^i_n + v^i \delta_{n,0} K, \label{repshift}\\
&& \tilde K = K, \\
&& \tilde D = D - v \cdot H_0. \label{zerograde}
\end{eqnarray}
These newly defined generators satisfy the commutation relation of
untwisted ones (\ref{extcartan}--\ref{halpha}); the twisted
algebra is isomorphic to the untwisted algebra (\ref{kk}). In the
Kaluza--Klein theory, where the bosonic description is based, the
charge and the mass are not distinguished. Thus, the ladder
operator (\ref{root}) carries mass $n+v \cdot \bar \alpha$.
Ultimately, we will be interested in the zero modes, or the
vanishing $\tilde D$ eigenvalues.

The weight of a twisted state is the eigenvalue of $\tilde H_0$
(Recall that $H_n$ with $n \ne 0$ is the ladder operator in the
$D$ direction.) This has an effect
\begin{equation}
 |p \rangle \to |\tilde p\rangle = |p + v\rangle.
\end{equation} Thus
\begin{equation}
L_0 |\tilde \Lambda \rangle = h_{\tilde \Lambda} |\tilde \Lambda
\rangle = h_{\tilde \Lambda} |\tilde p\rangle.
\end{equation}
So, we interpret that this is the representation of the twisted
algebra. In the $E_8 \times E_8$ theory, we cannot treat two $E_8$
groups separately in the twisted sector because of the relation
(\ref{repshift}). The mass shell condition for the highest
representation is
\begin{equation}
{M_{\rm L}^2 \over 4} = {(p+v)^2 \over 2} + \tilde N - \frac
    c{24}= h_{\tilde \Lambda} - \frac c{24}.
\end{equation}
Satisfied by the highest weight $\tilde \Lambda$, we have the same
explicit vector as (\ref{hgstwgtvec})
\begin{equation} \label{twhwv}
 p+v = \sum_j (A^{\h})^{-1}_{ij} \bar \alpha^{j \vee}.
\end{equation}
Again we will deal with the unbroken subgroup $\h$ arising from
breaking the original group $\g$, which is typically $E_8 \times
E_8$ or $SO(32)$ and in which $\h$ is embedded. Then here, we use
the inverse Cartan matrix $(A^{\h})^{-1}$ of the {\em subgroup $\h$}
and {\em original} $\g$ dual root $\bar \alpha^{j \vee}$. Now the
resulting $p$ does not necessarily belong to the untwisted zero mode
roots satisfying $p^2=2$, which is the case when the eigenvalue of
$\tilde D$, or $M^2_{\rm L} $ vanishes but not $D$. Also for the
state charged under semisimple and Abelian groups, the conformal
weight and weight vector are additive as in (\ref{sumh},\ref{sump})
with $p$ repleaced by $p+v$.

The twisted algebra depends on the shift vector $v$ only. The
argument is further extended to the $k$th twisted sector, in which
the only change is the effective shift vector $kv$. We also have
the symmetry $v \to -v$, meaning that we always have the
antiparticle($-v$) which has the complex conjugate representation
from that of the particle($v$). The chirality comes from right
movers which have only one helicity by Gliozzi--Scherk--Olive
(GSO) projection. A complete {\em chiral} state consists of {\em both}.

\subsection{Fixed point algebra}
Consider the invariant subalgebra under shifting (\ref{shift}),
whose elements satisfy
\begin{equation}
p \cdot v \in \Z.\label{intcondition}
\end{equation}
This is known as {\it fixed point algebra}and will be unbroken
subalgebra under orbifolding.

Given $v$ of finite order, there are only a few such fixed point
algebras possible \cite{CHK}. We can easily observe this from the
Dynkin diagram. Expressing the shift vector in terms of the
fundamental weight
\begin{equation} \label{shiftfund}
 v = \frac1l \sum_{i=1}^r s_i \bar \Lambda_i.
\end{equation}
We can show that the following is always satisfied \cite{FuSc}
\begin{equation} \label{kacpeterson}
 l = s_0 + \sum_{i=1}^r a_i s_i, \quad s_i \in \Z_{\ge 0},
\end{equation}
with the dual Coxeter label $a_i$ previously defined. By the other way
around, from a given shift vector we can always find an equivalent
one of this form. This `dominant' form is most convenient because we
can track the group theoretical origin of the action. We see that if
\begin{equation}
 {s_i \over l} = \bar \alpha^{i \vee} \cdot v
\end{equation}
is nonzero, then the corresponding root is a broken root as can be
seen from the relation (\ref{intcondition}). The nonzero integral
value also passes the condition, but this is not the case because the
above restriction implies $s_i<l$. Note the Cartan generators are
untouched and will provide the $U(1)$ generators.
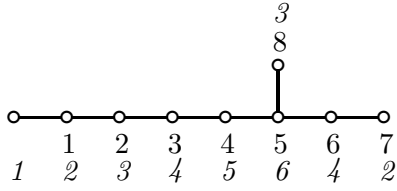
\begin{figure}[t] \label{fig:coxeter}
\begin{center}
\begin{picture}(150,40)
\thicklines \multiput(5,5)(20,0){7}{\circle{4}}
\multiput(7,5)(20,0){6}{\line(1,0){16}} \put(85,25){\circle{4}}
\put(85,7){\line(0,1){16}}
\multiputlist(6,-5)(20,0){1,2,3,4,5,6,7} {\it
\multiputlist(-14,-15)(20,0){1,2,3,4,5,6,4,2} \put(83,41){3}}
\put(83,30){8} \put(-15,5){\circle{4}}
 \put(-13,5){\line(1,0){16}}
\end{picture}
\end{center}
\caption{The extended $E_8$ Dynkin diagram and the ordering. The
italic numbers are the dual Coxeter labels.}
\end{figure}

Only a few set of integers $\{s_0,s_i\}$ satisfy the relation
(\ref{kacpeterson}) by the given order $l$, since $a_i$'s are positive.
To find the unbroken (fixed point) algebra, we only need to find a set of
nonnegative integers whose sum is order $l$ and then delete the circles
corresponding to the nonzero element. We can extend this argument
for more than one shift vector \cite{CHK}. These extra shift
vector(s) are provided by Wilson lines.

\section{Abelian charge}

Recall that under the shift (\ref{shift}) the Cartan generators
remain invariant, even when their roots are prevented by the
condition (\ref{intcondition}). This means that in the fixed point
algebra, they play the role of $U(1)$ generators and the rank is
preserved. The corresponding charge generator $q_i$ that projects
the state vector to give $U(1)$ charge $Q_i$
\begin{equation}
 Q_i = q_i\cdot p,
\end{equation}
or $\tilde p$ instead of $p$ in the twisted sector. We use the
same index $i$ since we have one to one correspondence between
$\bar \Lambda_i$ and $U(1)$ subgroups. The Abelian generators are
proportional to the fundamental weights used (corresponding to
$s_i \ne 0$) in the shift vector (\ref{shiftfund}),
\begin{equation}
 q_i \propto \bar \Lambda_i
\end{equation}
if the extended root of the original algebra is projected out $s_0
\ne 0$.  This is true because $q_i$ should be orthogonal to the
rest of the (simple) roots, otherwise this vector would be the
root vector of the corresponding nonabelian group. If the extended
root survives $s_0 = 0$, we can always find the following Abelian
generators $q$ (as many as the number of Abelian groups in the
fixed point group). By making linear combinations between the
fundamental weights used in the shift vector (\ref{shiftfund}),
allowing the negative coefficient of $s'_i$ we have
\begin{equation}
 q \propto \sum s'_i \bar \Lambda_i, \quad s'_i \in \Z
\end{equation}
satisfying
$$q \cdot \bar \theta = 0,$$
for it should be orthogonal to extended root $-\bar \theta$ of
the original algebra.

The normalization of $q_i$'s, related to the level $k$ and
determined by normalization of the current $T^a(z)$ \cite{FIQS}.
The corresponding vertex operator in this direction is $q_i \cdot
\partial_z X$, and has a different coefficient from (\ref{u1vertex}).
 From
(\ref{ope}), by fixing normalization of $f^{abc}$, as
(\ref{normalization}), the relative normalization of the $z^{-2}$ term
should be $k=q_i^2$ in this direction. For Abelian groups, the
structure constants vanish, and the normalization has to be fixed in
another way. However, at the compactification scale of an orbifold,
this $U(1)$ generator is embedded in $E_8\times E_8$ groups and
thus has definite normalization
\begin{equation}
 q_i^2 = k \label{abelnormal}
\end{equation}
to 1, as discussed before. The conformal weight for a state is
\begin{equation}
 h_{Q_i} = \frac12 {Q_i}^2 = \frac12 (q_i \cdot p)^2.
 \label{u1weight}
\end{equation}
Comparing to the similar relation (\ref{length}), we can determine a
$U(1)$ charged piece of vector $p$. Interestingly, it is also
proportional to $q_i$: The other parts of $p$ are fundamental
weights of the unbroken nonabelian group, which should not be
charged under this $U(1)$,
\begin{equation}
 q_i \cdot p = q_i \cdot r, \quad r \propto \bar \Lambda_i \propto q_i .
\end{equation}
This means that we can decompose the shift vector into completely
disconnected parts. The resulting state vector is
\begin{equation}
 p = \sum A^{-1}_{ij} \alpha^{j \vee} + r.
\end{equation}
The normalization of $q$ is fixed by (\ref{u1weight}). In
general, states may be charged under more than one $U(1)$'s: then
the vector is simply the addition of each $U(1)$ part.

There are potential anomalous $U(1)$'s. Since all the $U(1)$
generators belong to the original SO(32) or $E_8 \times E_8$, by
redefinition we can always absorb anomalies into one $U(1)$. This is
cancelled by the Green--Schwarz (GS) mechanism, and the charges of
the whole spectrum satisfy a specific `universality' condition. It
also fixes normalization \cite{CKM,FIQS,KN}, and our normalization
gives the correct answer. The statement of Ref. \cite{Gi} is for all
theory if we have at least one anomalous $U(1)$, the GS mechanism
fixes the normalization in four dimensional theory, regardless of
the origin of group breaking, which in this case is orbifolding.

We have stressed that the highest weight vector is the sum of the highest
weight vectors of disconnected semisimple parts. This also holds
true for the Abelian group, where now we have $c = 1/k=1$, the
`rank' of the Abelian group, and this is natural from the relation
(\ref{confanomaly}) with $d=1$ and $g=0$. It follows that the
conformal weight of a given state is the sum of conformal weights of
each simple or Abelian group.

\section{Orbifolding  spacetime}

\subsection{Compactification on orbifold}

To obtain a realistic theory, we compactify the string on an {\em
orbifold}. Practically, we define the orbifold as a torus modded by
a finite order automorphism $T^n / \bar{\sf P}$. Let $n$ be even and
pair the coordinates to complexify $Z^i = 2^{-1/2} (X^{2i-1}+i
X^{2i})$. We define the twist $\theta$ of $\bar{\sf P}$ by a
rotation on the diagonal entries of the spacetime group $SO(8)$ (the
massless little group of Lorentz $SO(1,9)$)
\begin{equation}
 \theta Z^i = \exp (2\pi i \phi_i) Z^i \label{spctwist}
\end{equation}
up to lattice translations. Because the action of $\bar{\sf P}$ is
a finite order and defined on the torus, it is known that there are
only thirteen kinds of lattice and twisting for {\it at least $N=1$
supersymmetry} \cite{DHVW}.

In this orbifold theory we have the twisted sector, since we have
closed strings modulo $\theta$. There is another condition
\begin{equation}
v^2 - \phi^2 =0, \quad {\rm mod~} 2/l. \label{modinv}
\end{equation}
They are required by the modular invariance of the string loop
amplitude \cite{DHVW,FV}.

The breaking of the gauge group occurs when we associate the
orbifold twist $\theta$ and the shift vector $v$. From (\ref{root})
we see that the grade (equivalently the mass squared, or eigenvalue
of worldsheet Hamiltonian) of the raising operator is changed by
$\bar \alpha \cdot v$ because of shifting (\ref{shift}). Physically,
the zero mode of this operator, corresponding to roots of gauge
symmetry, should commute with the Hamiltonian. Equivalently, the
state $| p \rangle$ should be invariant under (\ref{shift}). So the
resulting unbroken algebra is a fixed point algebra, obeying the
condition (\ref{intcondition}).

The matter spectrum is totally determined by the mass shell
condition (\ref{mommass}), supplemented by the generalized GSO
projection below. All of the matter spectrum forms the highest
weight module of the fixed point algebra. In the untwisted sector,
the matter representation is decomposed according to the
transformation property of $p \cdot v$. In each twisted sector, for
each highest weight representation
\begin{equation}
 \sum h_{\Lambda_i} - \frac c{24} = 0. \label{trace}
\end{equation}
where $c=c^{\g}+c^{\sf s}$ is the total conformal anomaly of the
gauge group (\ref{confanomaly}). For the spacetime degrees of
freedom in the $k$th twisted sector, it is given by zeta function
regularization,
\begin{equation}
 -\frac1{24} c^{\sf s} = \sum_i \sum_{n=1}^{\infty} (n+k\phi_i)
 = \sum_i \left(-\frac1{24} + \frac14 k\phi_i (1-k\phi_i)\right),
 \label{spcanomaly}
\end{equation}
where $i$ runs over the real spacetime bosonic degrees of freedom.
The sign is opposite for fermions. We adjusted $k\phi$ modulo
integer to lie in $0<k \phi_i<1$. By definition, the total zero
point energy is in $0 < c/24 < 1$. Inspecting the metric tensor of
weights, one can see that there are only a few states with
$h<\frac12$, so that only a few can be simultaneous representations
under more than one group.

In general, the orbifold action is not free, that is, we
have fixed points. The number of such localized spectra is that of
the fixed points $\chi$, from the Lefschetz fixed point theorem
\cite{DHVW},
\begin{equation}
 \chi = \det (1-\theta)=\prod 4 \sin^2 (\pi \phi_i),
 \label{numberfixedpt}
\end{equation}
over the compact dimension. Here, $\theta$ is regarded as the
rotation matrix of the lattice and it has integral elements. The
twisted string center of mass cannot have momentum and is localized
around fixed points.

However, when the order of the orbifold is non-prime, this naive $\chi$
does not work. By modular transformations some sectors can mix. By
successive shifting, a fixed point does not remain at that fixed
point any longer. For example, in the $T^6/\Z_4$ orbifold, all the
fixed points under the shift $\phi=\frac14(2~1~1~0)$ are not fixed
points under the shift $2\phi=\frac12(2~1~1~0)$. Here, we have a
nontrivial fixed representation. The remedy is to count the number of
effective fixed points (or tori). This can be read off by
integrating out the partition function. The projection
\begin{equation} \label{rightprojector}
 (\Delta_m)^n = \exp \left [ 2 \pi i n (\tilde N
 - N+(p+mv)\cdot v-(s+m\phi)\cdot \phi -\frac12(mv^2-m\phi^2) ) \right],
\end{equation}
projects out the invariant states under the orbifold action in the
$(\theta^m,\theta^n)$ twisted states \cite{IMNQ,FIQS}. (Also the
vector $p$ should belong to the $E_8\times E_8$ or $SO(32)$
lattice.) Thus, the number of effective fixed points in the
$\theta^m$th twisted sector is
\begin{equation}
P_m = \frac 1 l \sum_{n=0}^{l-1} \tilde \chi_{mn}
   (\Delta_m)^n. \label{proj}
\end{equation}
The new $\tilde \chi_{mn}$ is slightly different from the number of
fixed point $\chi$ because of the spin structure \cite{IMNQ}, but
can be fixed by modular invariance. This is a heterotic string
version of the GSO projection condition.

\subsection{Spacetime, or Lorentz symmetry}

Now consider states carrying the spacetime index. Here we focus on
the massless states, which are provided by worldsheet fermions that
transform under the Lorentz symmetry $SO(8)$. For the full
description of a supersymmetric right mover, we need supersymmetric
conformal algebra. In the Neveu--Schwarz (NS) sector, the lowest
lying state transforms as vector ${\bf 8}$. The Ramond (R) sector
has the lowest lying state, transforming as spinorial ${\bf 8}$.
When we bosonize them with four bosons, we have a unified
description. Similar to the weight vector of the gauge group $p$,
the state vector is denoted by $s$. The NS vector has the component
$(\pm 1~0~0~0)$ up to permutations and the R spinor has
$(\pm\frac12~\pm\frac12~\pm\frac12~\pm\frac12)$ with an even number
of minus signs. These are all massless. Not only these massless
states, but all the excited states form the representation of the
$SO(8)$ affine Lie algebra.

We can easily examine the symmetry breaking of $SO(8)$ using the
fixed point algebra as discussed before. For example, the $T^6/\Z_3$
twist is given by $\phi=\frac13(2~1~1~0)=\frac13(\bar \Lambda_1+\bar
\Lambda_3+\bar \Lambda_4)$. (We are using the fundamental weights of
$SO(8)$.) Deleting the corresponding nodes from the extended Dynkin
diagram, one notes that the surviving group is
$$SU(3)\times U(1)^2.$$
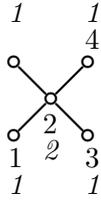
\begin{figure}[t] \label{fig:so8}
\begin{center}
\begin{picture}(50,40)
\thicklines \multiput(5,0)(14,14){3}{\circle{4}}
\put(5,28){\circle{4}} \put(33,0){\circle{4}}
\multiput(7,2)(14,14){2}{\line(1,1){10}}
\multiput(7,26)(14,-14){2}{\line(1,-1){10}} \put(16,1){2}
\put(3,-12){1}  \put(32,-12){3} \put(32,33){4}
 {\it \put(16,-9){2}
\put(3,-22){1} \put(3,43){1} \put(32,-22){1} \put(32,43){1}}
\end{picture}
\end{center}
\caption{The extended $SO(8)$ Dynkin diagram. The italic numbers
are the dual Coxeter labels.}
\end{figure}
Accordingly, this leads to branching $\bf 8= 3+\bar 3 +1 +1$, which
can be checked by the transformation property of $s \cdot \phi$. The
$SU(3)$ is the holonomy group and the remaining $U(1)$'s are the $R$
symmetry and the noncompact $SO(2)$ Lorentz symmetry in the light
cone gauge. When we compactify on a four dimensional orbifold
$T^4/\Z_l$, with shift $\frac1l(1~1~0~0)=\frac1l\bar \Lambda_2$ the
unbroken Lorentz symmetry $SO(4)=SU(2)\times SU(2)$ is not
explicitly manifest.

The same argument applies to the twisted sector. The orbifold action
(\ref{spctwist}) is basically the same as shifting (\ref{shift}).
Thus under the twisting, they are shifted as $s \to s+\phi$,
satisfying the relation (\ref{repshift}). The bosonic degrees of
freedom have the same Lorenz symmetry. Since we know every inner
automorphism can be converted into a shift vector, without knowing
the concrete representation, we have a corresponding conformal weight
$h_\Lambda$. Alternatively, we may understand the spacetime
contribution in terms of an oscillator. Through orbifolding, now we
have a shifted oscillator mode number by $\phi_i$ affecting a mass
shell condition such as $h_\Lambda$; however then the algebraic
property becomes less transparent.

Note that, to the bosonic degrees of freedom, the spacetime twisting
(\ref{spctwist}) does touch the Cartan generator itself. Care must
be taken in dealing with them. Using the fact that every finite
inner automorphism can be representation, the spacetime twisting can
be converted to twisting with shift vector, which have a suitable
basis.

We may use this fact for orbifolding and obtaining the massless
spectrum in other superstring theories.

\section{Examples}

\subsection{Standard embedding}

Take an example of $E_8 \times E_8$ theory compactified on
$T^6/\Z_3$ orbifold with twist $\phi=\frac13(2~1~1~0)=\frac13(\bar
\Lambda_1+\bar \Lambda_3+\bar \Lambda_4)$ in the $SO(8)$ basis.
Deleting the corresponding nodes in the Dynkin diagram, the
resulting spacetime degrees of freedom are $SU(3)\times U(1)
\times SO(2)$ as before. For the standard embedding of the group
degrees of freedom, $v=\frac13(2~1~1~0^5;0^8)=\frac13(\bar
\Lambda_2;0)$ in the $E_8$ basis, we obtain an unbroken group as
$$ SU(3) \times E_6 \times E_8. $$
To find the matter spectrum, we need only check a few. From the
integrability condition (\ref{integrability}) the candidates are,
$$
 \textstyle h_{\bf 3} = \frac13,\quad h_{\bf 27} = \frac23,
 \quad h_{\bf (3,27)}=h_{\3}+h_{\bf 27}=1
$$
The antiparticles have the same conformal weights. The GSO
projection condition determines which one survives. Under the
spacetime and the gauge group, $(SU(3);SU(3)\times E_6;E_ 8)$, the
untwisted sector has
$$ 3 (1\bf;3,27;1) $$
The multiplicity 3 came from the right movers. Numbers are in
boldface, except those for spacetime representation.

In the twisted sector, we have the zero point energy $-\frac
c{24}= -\frac1{24} (c^\g+c^{\sf s})=-\frac23 $ from
eq.(\ref{spcanomaly}). The $\bar \Lambda_1=\bf 27$ of $E_6$ has
conformal weight $\frac23$. The $\bar \Lambda_2=\bar{\bf 3}$ of
$SU(3)$ has $\frac13$, regardless of whether they come from
the Lorentz or gauge group. So the combination of $(3;\bf \bar 3,1;1)$
under the Lorentz group $SU(3)$ and gauge group $SU(3)$ also has
a total conformal weight of $\frac13+\frac13=\frac23$. The
corresponding highest weight vectors $\tilde p$ are
\begin{eqnarray}
 ({\bf 1, 27;1})  &:& \sum_j (A^{E_6})^{-1}_{1j} \bar \alpha^j =  \textstyle
(\frac23~-\frac23~-\frac23~0^5;0^8 ) \notag \\
 ({\bf \overline 3,1;1})  &:& \sum_j (A^{SU_3})^{-1}_{2j} \bar \alpha^j
  = \textstyle (-\frac13~~\frac13~-\frac23~0^5;0^8).
 \notag
\end{eqnarray}
The $\bar \alpha^j=\bar \alpha^{j \vee}$ is a simple root in the
$E_8\times E_8$ basis and each inverse Cartan matrix
$(A^{\h})^{-1}$ is of the simple subgroups $E_6$ and $SU(3)$. Here
we suppressed the eigenvalue of $(D,K)$. Note that the vector
$p=\tilde p-v$ belongs to $E_8 \times E_8$ {\em lattice}, not
necessarily being a root with $p^2=2$.

With multiplicity from the number of fixed points $\chi=27$ from
eq.(\ref{numberfixedpt}), we have
$$
 27(3;\overline{\3},\1;\1)+27(1;\1,{\bf 27};\1),
$$
which survives under the projection (\ref{proj}). The antiparticles
come from the second twisted sector with twist $2\phi$, or
equivalently $-\phi$.

\subsection{Models having an Abelian group}

Consider again the $T^6/\Z_3$ example with the shift vector
$v=\frac13(2~0^7;1~1~0^6)=\frac13 (\bar \Lambda_7 ;\bar
\Lambda_1)$. We can check that the modular invariance condition is
satisfied and the resulting gauge group is $SO(14)\times U(1)
\times E_7 \times U(1)$. The two $U(1)$ generators are $q_7 =
\frac12(\bar \Lambda_7;0)$ and $q_1' =\frac1{\sqrt2}(0;\bar
\Lambda_1)$ by the normalization (\ref{abelnormal}). Note that
this gives correct normalization \cite{KN} for the GS mechanism.

In view of the branching rule, in the untwisted sector we obtain
$$ 3 (1{\bf;14;1)}+3 (1{\bf;64;1)}+3 (1{\bf;56;1)}+ 3
(1{\bf;1;1)}.
$$
In the twisted sector, the zero point energy is still $-\frac
c{24}=-\frac23$. The $SO(14)$ vector with $h_{\bf 14}=\frac12$ alone
cannot be massless, but should have other components to fulfill the
mass shell condition. The missing mass is provided by other vectors
$r_7$ and $r_1'$ charged under $U(1)$'s.

The corresponding highest weight vector has the form
$$ \tilde p = \sum_j (A^{SO(14)})^{-1}_{1j}\bar \alpha^j + r_7 +r_1'.$$
The first term is $\bar \Lambda_1$ of $SO(14)$. The $r_7$ and $r_1'$
are also proportional to $(\bar \Lambda_7;0)$ and $(0;\bar
\Lambda_1)$, respectively. They are completely fixed by the
condition
$$ h_Q =
\frac12 (q_7 \cdot r_7)^2 + \frac12 (q_1' \cdot r_1')^2=
\frac16,$$ and the generalized GSO projection condition. The
resulting vector is
$$ \textstyle \tilde p = (0~1~0^6;0^8)+(-\frac13)(1~0^7;0^8) +\frac13(0^8;1~1~0^6), $$
and charged as $(1;{\bf 14;1})$. The Lorentz $\bf 3$ of $SU(3)$ can
contribute $h=\frac13$ and it provides another charged state,
$(3;\bf 1;1)$. In addition, there is a state which is a singlet
under the whole nonabelian group $(1;\bf 1;1)$. They all have
multiplicity $\chi = 27$.

\subsection{Non-prime orbifold}

We take the $T^6/\Z_4$ orbifold example with $\phi=\frac14(2~1~1~0)$
and the standard embedding in the group space. The resulting
spacetime symmetry is $SU(2) \times U(1)^2$ and the group degree of
freedom is $E_6\times SU(2)\times U(1)\times E_8$. The $U(1)$
generator is $q_2=\frac1{\sqrt6}(\bar \Lambda_2;0)$. In the
untwisted sector, by the branching rule we have
$$ 2 (1;{\bf 27,2;1})+2(1;{\bf \overline{27},1;1})
+2(1;{\bf 27,1;1})+2(1;{\bf 1,2;1}) $$ with the multiplicity 2
from the spacetime $SU(2)$ doublet right movers.

In the first twisted sector the zero point energy is
$-\frac{11}{16}$. We have $h_{\bf 27}=\frac23,\ h_{\bf 2}=\frac12$
which determines the nonabelian part of the state only. The missing
mass is from the $U(1)$ charge. By the same argument, the weight
vector have the $U(1)$ parts $q$ as $\frac1{12}(\bar \Lambda_2;0),\
\frac14(\bar \Lambda_2;0)$ respectively.

Let us see the second twisted sector whose effective shift is
$2\phi=\frac12(2~1~1)$. Its zero point energy is $-\frac34$. We
expect a vectorlike spectrum since the $k$th twisted sector
spectrum is the same as that of the $(l-k)$th. The Abelian charge
for $\bf 27$ is $\frac16(\bar \Lambda_2;0)$.

Resorting to the projection condition (\ref{proj}), we have ten
$\bf 27$'s and six ${\bf \overline{27}}$'s. Also, we observe that
the modular invariance condition completely determines the $U(1)$
part of a given vector.

\section{Discussion}
We have seen that the spectra of heterotic strings on orbifolds can
be obtained from a simple relation between conformal weight and
mass. This relation is natural in view of conformal field theory.
The connection between conformal field theory and affine Lie algebra
is provided by Sugawara construction.

This proves to be useful when we are interested in the twisted
sector. By orbifolding with a shift vector, we associated the
spacetime point group action with the group degree of freedom.
This twisting mixes the massless and massive states, and they are
represented by twisted affine Lie algebra. In fact, they are
elements of another, independent Hilbert space of twisted states.
The conformal weights, hence the explicit states in the twisted
state, can be easily obtained because the twisted algebra is
isomorphic to untwisted one.

Also we stress that this is the only systematic way to obtain
Abelian generators. The corresponding conformal weights and
representation vectors are treated on an equal footing as those of
nonabelian cases. It is hoped that this will reveal the role of anomalous
$U(1)$.

We observe that the resulting spectrum is strongly related to
branching rules \cite{Ka}. For the purely group degree of freedom,
there are some mathematical discussions on branching rules of
affine Lie algebra. From the partition function (called a
character in algebra) we can relate representations between a
given algebra and an embedded subalgebra. However, in string
theory some characters are different. The low energy fields
acquire chirality due to right movers and the multiplicity comes
from the invariance property of states under the point group.
However this nature is reflected in the same way and also leads to
the same rule.

If such a unified branching rule is accessible, we are able to have
a better understanding of, for example, the anomaly freedom of
orbifold theory \cite{FV}: we know that a representation of
subalgebras embedded from an anomaly-free representation is also
anomaly-free. The modular invariance is crucial for an anomaly-free
theory. Using the branching function \cite{Ka} and anomaly
polynomial \cite{SW} it will be observed that the absence of anomaly
originates from the fact that there is no modular form of weight two.

We can apply this idea to other theories described on the lattice:
the Narain compactification and the free fermionic formulation just
by interpreting shift and twisting fields. Also it holds for
orbifold compactification of other theories, for example Type II
strings. The method we present in this paper will, we hope, provide
a guideline for a top-down approach of model building.

\acknowledgements{The author is grateful to Jihn E. Kim for his
valuable input. This work is supported in part by the KOSEF ABRL
Grant No. R14-2003-012-01001-0 and the BK21 program of the Korean
Ministry of Education.}

\appendix
\section{Algebraic conventions}
In this appendix, we define some algebraic elements used in the
examples, following \cite{FuSc} where the complete list is
available. The orthogonal representation for simple roots of $E_8$
are
$$
\begin{matrix}
\bar \alpha^1 = ( & 0 & 1 & -1 & 0 & 0 & 0 & 0 & 0 & ) \\
\bar \alpha^2 = ( & 0 & 0 & 1 & -1 & 0 & 0 & 0 & 0 & ) \\
\bar \alpha^3 = ( & 0 & 0 & 0 & 1 & -1 & 0 & 0 & 0 & ) \\
\bar \alpha^4 = ( & 0 & 0 & 0 & 0 & 1 & -1 & 0 & 0 & ) \\
\bar \alpha^5 = ( & 0 & 0 & 0 & 0 & 0 & 1 & -1 & 0 & ) \\
\bar \alpha^6 = ( & 0 & 0 & 0 & 0 & 0 & 0 & 1 & -1 & ) \\
\bar \alpha^7 = ( & \frac12 & -\frac12 & -\frac12 & -\frac12 & -\frac12 & -\frac12 & -\frac12 & \frac12 & ) \\
\bar \alpha^8 = ( & 0 & 0 & 0 & 0 & 0 & 0 & 1 & 1 & ).
\end{matrix}
$$
The extended root $\bar \alpha^0 = (-1~-1~0~0~0~0~0~0) = -\bar
\theta$ is defined as the negative of highest root. All the root
vectors are self-dual $\bar \alpha^i = \bar \alpha^{i \vee}$
because $\bar \alpha^2=2$. Accordingly, we have fundamental
weights,
$$
\begin{matrix}
\bar \Lambda_1 = ( & 1 & 1 & 0 & 0 & 0 & 0 & 0 & 0 & ) \\
\bar \Lambda_2 = ( & 2 & 1 & 1 & 0 & 0 & 0 & 0 & 0 & ) \\
\bar \Lambda_3 = ( & 3 & 1 & 1 & 1 & 0 & 0 & 0 & 0 & ) \\
\bar \Lambda_4 = ( & 4 & 1 & 1 & 1 & 1 & 0 & 0 & 0 & ) \\
\bar \Lambda_5 = ( & 5 & 1 & 1 & 1 & 1 & 1 & 0 & 0 & ) \\
\bar \Lambda_6 = ( & \frac52 & \frac12 & \frac12 & \frac12& \frac12& \frac12& \frac12& -\frac12 & ) \\
\bar \Lambda_7 = ( & 2 & 0 & 0 & 0 & 0 & 0 & 0 & 0 & ) \\
\bar \Lambda_8 = ( & \frac32 & \frac12 & \frac12& \frac12 &
\frac12 & \frac12& \frac12 & \frac12& ).
\end{matrix}
$$
They are used in the standard form of shift vector in
(\ref{shiftfund}). We have also used this method to break the
spacetime Lorentz group $SO(8)$, whose simple roots and
fundamental weights are
$$
\begin{matrix}
\bar \alpha^1 = ( & 1 & -1 & 0 & 0 & ) \\
\bar \alpha^2 = ( & 0 & 1 & -1 & 0 & ) \\
\bar \alpha^3 = ( & 0 & 0 & 1 & -1 & ) \\
\bar \alpha^4 = ( & 0 & 0 & 1 & 1 & ),
\end{matrix}
\qquad
\begin{matrix}
\bar \Lambda_1 = ( & 1 & 0 & 0 & 0 & ) \\
\bar \Lambda_2 = ( & 1& 1 & 0 & 0 & ) \\
\bar \Lambda_3 = ( & \frac12 & \frac12 & \frac12 & -\frac12 & ) \\
\bar \Lambda_4 = ( & \frac12 & \frac12 & \frac12 & \frac12  & ).
\end{matrix}
$$
Now we define the quadratic form matrices, defined as the inverses
of the Cartan matrices $A_{ij} = 2 a^i \cdot a^{j \vee}$. They are
used for finding the conformal weight in (\ref{length}) and
obtaining the corresponding highest weight vector in
(\ref{hgstwgtvec},\ref{twhwv}). We take examples of $SU(3), E_6$
and $E_8$ used in the standard embedding of $T^6/\Z_3$ orbifold,
$$
(A^{SU(3)})^{-1} = \frac13 \begin{pmatrix} 2 & 1 \\ 1 & 2
\end{pmatrix}, \quad
(A^{E_6})^{-1} =  \frac13
\begin{pmatrix}
4 & 5 & 6 & 4 & 2 & 3 \\
5 & 10 & 12 & 8 & 4 & 6 \\
6 & 12 & 18 & 12 & 6 & 9 \\
4 & 8 & 12 & 10 & 5 & 6 \\
2 & 4 & 6 & 5 & 4 & 3 \\
3 & 6 & 9 & 6 & 3 & 6\\
\end{pmatrix}
$$ $$
\quad (A^{E_8})^{-1} =
\begin{pmatrix}
2 & 3 & 4 & 5 & 6 & 4 & 2 & 3 \\
3 & 6 & 8 & 10 & 12 & 8 & 4 & 6 \\
4 & 8 & 12 & 15 & 18 & 12 & 6 & 9 \\
5 & 10 & 15 & 20 & 24 & 16 & 8 & 12 \\
6 & 12 & 18 & 24 & 30 & 20 & 10 & 15 \\
4 & 8 & 12 & 16 & 20 & 14 & 7 & 10 \\
2 & 4 & 6 & 8 & 10 & 7 & 4 & 5 \\
3 & 6 & 9 & 12 & 15 & 10 & 5 & 8 \\
\end{pmatrix}.
$$

\end{document}